\documentclass[10pt, letterpaper]{article}

\newcommand{\reff}[1]{(\ref{#1})}
\newcommand{\beq}{\begin{equation}} \newcommand{\eeq}{\end{equation}}
\newcommand{\beqa}{\begin{eqnarray}}    \newcommand{\eeqa}{\end{eqnarray}}
\newcommand{\btab}{\begin{tabular}}     \newcommand{\etab}{\end{tabular}}

\newcommand{\bt}{\begin{table}}     \newcommand{\et}{\end{table}}

\newcommand{\ba}{\begin{array}}     \newcommand{\ea}{\end{array}}
\newcommand{\bc}{\begin{center}}        \newcommand{\ec}{\end{center}}
\newcommand{\bfig}{\begin{figure}}      \newcommand{\efig}{\end{figure}}
\newcommand{\bp}{\begin{picture}}       \newcommand{\ep}{\end{picture}}
\newcommand{\bq}{\begin{quote}}     \newcommand{\eq}{\end{quote}}
\newcommand{\ben}{\begin{enumerate}}    \newcommand{\een}{\end{enumerate}}

\newcommand{\mc}[1]{\ensuremath{\mathcal{#1}}}

\newcommand{\sm}[1]{\ensuremath{m_{(#1)}}}

\newcommand{\mtext}[1]{\ensuremath{ \quad \mbox{#1}\quad } }
\let\La\Lambda

\begin{document}

\title{Supersymmetric Calogero-Moser-Sutherland models:
superintegrability structure and eigenfunctions
    }

\author{Patrick Desrosiers\thanks{pdesrosi@phy.ulaval.ca} \\
\emph{D\'epartement de Physique, de G\'enie Physique et d'Optique},\\
Universit\'e Laval, \\
Qu\'ebec, Canada, G1K 7P4.
\and
Luc Lapointe\thanks{lapointe@inst-mat.utalca.cl }\\
\emph{Instituto de Matem\'atica y F\'{\i}sica},\cr
Universidad de Talca,\cr
Casilla 747, Talca, Chile.
\and
Pierre Mathieu\thanks{pmathieu@phy.ulaval.ca} \\
\emph{D\'epartement de Physique, de G\'enie Physique et d'Optique},\\
Universit\'e Laval, \\
Qu\'ebec, Canada, G1K 7P4.
}

\date{October 2002}

\maketitle

\begin{abstract}

We first review the construction of the supersymmetric extension of
the quantum
Calogero-Moser-Sutherland (CMS) models. We stress the remarkable fact
that this extension is
completely captured by the insertion of a fermionic exchange operator
in the Hamiltonian: sCMS models ({\it s} for
supersymmetric) are nothing but special exchange-type CMS models.
Under the appropriate projection, the conserved charges can
thus be formulated in terms of the standard   Dunkl operators.  This
is illustrated in the rational case, where the
explicit form of the $4N$ ($N$ being the number of bosonic variables)
conserved charges is presented, together with their
full algebra.  The existence of
$2N$ commuting bosonic charges settles the question of the
integrability of the srCMS model.  We then prove its
superintegrability by displaying $2N-2$ extra independent charges
commuting with the Hamiltonian. In the second  part, we
consider the supersymmetric version of the trigonometric  case (stCMS
model) and review the construction of its
eigenfunctions, the Jack superpolynomials. This leads to
closed-form expressions, as  determinants of determinants
involving supermonomial symmetric functions. Here we focus on the main ideas
and the generic aspects of the
construction: those applicable to all models whether supersymmetric or not.
   Finally, the possible Lie superalgebraic structure
underlying the stCMS model and its eigenfunctions is briefly
considered.\footnote{To appear in the proceedings of the {\it
Workshop on superintegrability in classical and quantum systems},
September 16-22 2002, Centre de recherches math\'ematiques,
Universit\'e de Montr\'eal, ed. by P. Winternitz.}
\end{abstract}

\newpage


\section{Introduction}

This presentation has two main objectives. The first is to stress
the interpretation of the supersymmetric extension of the quantum
Calogero-Moser-Sutherland (CMS) models \cite{CMS}\footnote{We use
the qualitative `Calogero-Moser-Sutherland' to describe the
generic class of models that includes the models studied by
Calogero and Sutherland in the quantum case  and by Moser in the
classical case. In this work, however, we only treat the quantum
case. In this context the name of Moser is often omitted. The
rationale for its inclusion is due to the fundamental importance
of the Lax formulation in the quantum case, which is a direct
extension of the classical one that he introduced.}  as special
exchange-type CMS models \cite{Poly}.  This provides a direct path
for establishing their integrability \cite{DLM1} and also, as is
first shown here, their superintegrability. For this analysis,
which is the content of section 2, we focus on the rational case.
Our second goal is to provide a simple quasi-qualitative
presentation of the main ideas underlying our recent construction
of the supersymmetric trigonometric CMS (stCMS) eigenfunctions,
 that is,  of the Jack superpolynomials \cite{DLM1,DLM2,DLM3}.
This is the content of section 3. Most technicalities are avoided.
We emphasis that our method is a general program for constructing
explicitly the eigenfunctions of a CMS-type Hamiltonian,
supersymmetric or not. Some subsections  are thus formulated in
rather general terms.

\section{Integrability and superintegrability of the srCMS model}

\subsection{Supersymmetric quantum mechanics and fermionic-exchange operators }

Consider a quantum model, whose Hamiltonian is denoted  $\mc{H}$,
that contains bosonic and fermionic variables.  That is,
that contains in
addition to the $2N$ bosonic variables $(x_i,p_i)$, the $2N$ fermionic
variables  $(\theta_i,\theta_i^\dagger)$ (e.g.,
$\theta_i\theta_j=-\theta_j\theta_i$, so that $\theta_i^2=0$).
   The bosonic
and fermionic variables
satisfy respectively a Heisenberg and a Clifford
algebra:
\beq
   [x_j,p_k]=i \delta_{jk} \; , \qquad\quad\{\theta_j,
\theta_k^\dagger\}=\delta_{jk}, \label{algquant} \eeq where
$\{\;,\;\}$ stands for the anticommutaror.  All other commutators
or anticommutators are equal to zero.  We usually work with a
differential realization of these algebras: \beq p_j=-i
\frac{\partial}{\partial x_j} \; ,\qquad\qquad
\theta_j^\dagger=\frac{\partial}{\partial \theta_j}.
\label{diffrepre} \eeq

To construct a supersymmetric Hamiltonian, we will first construct  two
supersymmetric charges, denoted
$Q$ and $Q^\dagger$, and define the
   Hamiltonian  as their anticommutator:
\beq
      \mc{H}=\frac{1}{2}\{Q, Q^\dagger\} \, .
\eeq
The Hamiltonian is invariant under a supersymmetric transformation if:
\beq
      Q^2=(Q^\dagger)^2=0. \label{consusyq}
\eeq
   With the charges written under the form
   \beq Q=\sum_{i=1}^N
\theta^\dagger_i A_i(x,p) ,\quad Q^\dagger =\sum_{i=1}^N
\theta_i A^\dagger_i(x,p)\, ,
   \eeq
   eq. \reff{consusyq} requires:
\beq \label{consusyqq}
[A_i,A_j]=0=[A_i^\dagger,A_j^\dagger] ,\quad\forall \, i,j. \eeq
The generic supersymmetric Hamiltonian thus reads:
\beq
\mc{H}=\frac{1}{2}\left(\sum_i A_i^\dagger A_i
+\sum_{i,j}\theta_i^\dagger\theta_j [A_i,A_j^\dagger] \right)\, .
\label{defhgen}
\eeq
   For non-relativistic models, the Hamiltonian
is proportional to the square of the particles' momenta.  This forces $A_i$
to be a linear function of the momentum $p_i$, that is: \beq \label{defQ}
Q=\sum_j \theta_j^\dagger(p_j-i\Phi_j(x)), \quad Q^\dagger=\sum_j
\theta_j(p_j+i\Phi_j(x)) \, .
  \label{constcharge} \eeq Condition \reff{consusyqq}
imposes that the potential $\Phi_j(x)$ be of the form
\beq \label{prepot}
\Phi_j(x)=\partial_{x_j}W(x)\,,
\eeq
   where $W(x)$
(called the prepotential), is an arbitrary
function of the variables $x_1,\dots,x_N$. The supersymmetric
Hamiltonian now takes the form
   \cite{Free}:
\beq
\mc{H}=\frac{1}{2}\sum_{i}(p_i^2+(\partial_{x_i}W)^2+\partial^2_{x_i}W)-\sum_{i,j}\theta_i\theta_j^\dagger\partial_{x_i}\partial_{x_j}W\,
. \eeq This Hamiltonian is an extension of the purely bosonic
model whose potential is
   $\sum_i[(\partial_{x_i}W)^2+\partial^2_{x_i}W]$.  We stress that
this construction fixes uniquely the supersymmetric
extension of the model to be considered.
Specializing to a  symmetric prepotential that can be broken up
into a sum of two-body interactions: \beq
W(x)=\sum_{i<j}w(x_{ij})\, \qquad{\rm with} \qquad
   w'(x_{ij})=X_{ij} \, , \eeq  where $X_{ij}$ stands for an antisymmetric
function of $x_{ij}=x_i-x_j$,  our Hamiltonian
reads:
  \beq
\mc{H}=\frac{1}{2}\sum_{i} p_i^2 +\sum_{i <
j}[X_{ij}^2+X'_{ij}(1-\theta_{ij} \theta^\dagger_{ij})]+
\sum_{i<j<k}Y_{ijk}\, , \eeq with \beq
Y_{ijk}=X_{ij}X_{ik}+X_{ji}X_{jk}+X_{ki}X_{kj}\, .
   \eeq
This is the supersymmetric Hamiltonian we were looking for.

The main observation at this point  is the following:
   the term
\beq
\kappa_{ij}\equiv 1-\theta_{ij}\theta^\dagger_{ij}=
1-(\theta_{i}-\theta_j)(\partial_{\theta_i}-\partial_{\theta_j}).
\eeq
which captures the whole dependence of the Hamiltonian upon the
fermionic variables, is a fermionic-exchange
operator \cite{SriramShastry:1993cz}.  In other words, its action on
an arbitrary function  $g$ of the fermionic variables
$\theta_i$ and $\theta_i^\dagger$ reads
\beq
\kappa_{ij}\,
g(\theta_i,\theta_j,\theta^\dagger_i,\theta^\dagger_j)=
g(\theta_j,\theta_i,\theta^\dagger_j,\theta^\dagger_i)\, \kappa_{ij}\, .
\eeq
In addition, it satisfies all the properties
   of an exchange operator:
\begin{equation}
\kappa_{ij}= \kappa_{ji},\quad
       \kappa_{ij}^\dagger= \kappa_{ij},\quad
       \kappa_{ij} \kappa_{jk}= \kappa_{ik} \kappa_{ij}= \kappa_{jk}
\kappa_{ki},\quad
       \kappa_{ij}^2=1\, .
\end{equation}
   That the supersymmetric extension is fully captured by the
introduction of a fermionic exchange
operator is thus a generic feature of supersymmetric many-body
problems whose interaction is decomposable into a sum of two-body
interactions. This is also a key technical tool in our
subsequent analysis, as we will shortly explain.  But we would like
to make the observation
that,  up this point, the discussion is quite general and {\it not
restricted to integrable problems}.

Having an Hamiltonian expressed in terms of a fermionic exchange
operator implies that under the right projection, we can
trade $\kappa_{ij}$ for the ordinary exchange operator,
$K_{ij}$, that exchanges the variables $x_i$ and $x_j$:
\beq
          K_{ij}f(x_i, x_j)=f(x_j, x_i,)K_{ij} \, ,
\eeq
with  $f(x_i,x_j)$ standing for a function or
an operator.
The suitable projection is in our case the one on the space $P^{S_N}$ of
functions invariant under the simultaneous
exchange of the bosonic and the fermionic variables, that is,
invariant under the action of
\beq
{\cal K}_{ij}= K_{ij}\kappa_{ij} \, .
\eeq
On this space, the supersymmetric Hamiltonain reduces to an ordinary
exchange-type Hamiltonian \cite{DLM1}. Stated differently:
acting on the proper space, we can study many features of the supersymmetric
model {\it without even introducing fermionic
degrees of freedom!}

\subsection{The integrability of the srCMS model }
To work with a concrete example,  consider the simple  srCMS model \cite{Free}:
\begin{equation}
\mc{H}_{(r)}=-\frac{1}{2}\sum_{j=1}^N \partial_{x_j}^2+\sum_{1\leq j <
k\leq N}\frac{\beta(\beta-\kappa_{jk})}{(x_j-x_k)^2}.
\end{equation}
We will  establish the integrability of this model using the
Dunkl-operator formalism, thus relying heavily on the
projection trick.

Using the rational Dunkl operators,
\begin{equation}
D_j=\partial_{x_j}-\beta\sum_{k\neq j}\frac{1}{x_{jk}}K_{jk} \, ,
\end{equation}
we find
\begin{equation}
-\frac{1}{2}\sum_{j=1}^N D_j^2 =
-\frac{1}{2}\sum_{j=1}^N \partial_{x_j}^2+\sum_{1\leq j <
k\leq N}\frac{\beta(\beta-K_{jk})}{(x_j-x_k)^2} \, .
\end{equation}
The Dunkl operators are commuting and covariant:
\begin{equation}
[D_j,D_i]=0\;\;,\qquad K_{ij}D_j=D_iK_{ij} \, .
\end{equation}
Therefore, the conservation laws of the rCMS model (with exchange
terms) are simply
   $\sum_{j}D_j^n $, for $ n=1,\ldots,N$.
The conserved charges of the rCMS model without exchange terms are obtained
by simply projecting these expressions onto the
space of symmetric functions, which amounts to replacing every factor
$K_{ij}$ by 1 once pushed to the right.  On the other
hand, the conserved charges of the srCMS model are obtained as follows:
\beq
\mc{H}_n=\sum_{j}D_j^n  \,  \Big|_{P^{S_N}}\, , \quad n=1,\ldots,N \, .
\eeq
  In particular, $\mathcal{H}_2=-2\mathcal{H}_{(r)}$.
The
explicit dependence of the srCMS conserved charges
$\mc{H}_n$ upon the fermionic variables can be obtained by
implementing the projection, that is, replacing every factor
$\kappa_{ij}$ by
$K_{ij}$ once shifted to the right.   The proof
of the commutativity of these charges
leans on a simple property of the projection: \beq
\left[A\,\Big|_{P^{S_N}},B\,\Big|_{P^{S_N}}\right]=
[A,B]\,\Big|_{P^{S_N}}\qquad \mbox{ if
}\qquad [\mc{K}_{ij},A]=[\mc{K}_{ij},B]=0 \label{propprojec} \, . \eeq The
operators $\mc{H}_{n}$ meet this requirement since $
[\mc{K}_{ij},(\sum_k {D}_k^n)]=0$.  Therefore, the commutator can be
evaluated before doing the projection and its vanishing
is an immediate consequence of the commutativity of the Dunkl operators.

We have thus found the supersymmetric extension of all the usual rCMS
charges and proved their commutativity. This
however does not imply the integrability of the srCMS model because
there are more degrees of freedom in the supersymmetric
case (these are the $2N$ extra Grasmannian variables). Consequently,
one should expect more conserved charges.  It is
actually rather easy to construct $3N$ extra charges involving
explicitly (i.e., even before the projection) some fermionic
variables
\cite{DLM1}:
\beq
\ba{rcl}
\mc{Q}_{(n)}&=&\sum_i \theta_i D_i^n \,
\Big|_{P^{S_N}}\,\phantom{\theta^\dagger}\mbox{ , }n=0,1,\ldots,
N-1\, , \\
\mc{Q}_{(n)}^\dagger&=&\sum_i \theta_i^\dagger D_i^n\,
\Big|_{P^{S_N}}\,\phantom{\theta^\dagger} \mbox{ , }n=0,1,\ldots,
N-1\, ,\\
\mc{I}_{(n)}&=&\sum_i
\theta_i\theta_i^\dagger D_i^n \,  \Big|_{P^{S_N}}\,\mbox{ , }n=0, 1,
\ldots, N-1 \, .
\ea
\eeq
These charges satisfy the following algebra:
\beq
\ba{c}
\{\mc{Q}_{(n)}, \mc{Q}_{(m)}\}=
\{\mc{Q}_{(n)}^\dagger,\mc{Q}_{(m)}^\dagger\}=
\left[\mc{I}_{(n)},\mc{I}_{(m)}\right]=0\,
, \\
\left[\mc{Q}_{(n)},\mc{H}_{(m)}\right]=
\left[\mc{Q}^\dagger_{(n)},\mc{H}_{(m)}\right]=
\left[\mc{I}_{(n)},\mc{H}_{(m)}\right]=
\left[\mc{H}_{(n)},\mc{H}_{(m)}\right]=0\,
.
\ea \eeq
together with
\beq
\ba{c}
\{\mc{Q}_{(n)}, \mc{Q}_{(m)}^\dagger\}=
\mc{H}_{(n+m)}\, , \\
\left[\mc{Q}^\dagger_{(n)},\mc{I}_{(m)}\right]=
\mc{Q}^\dagger_{(n+m)}
\mtext{,}\left[\mc{Q}_{(n)},\mc{I}_{(m)}\right]=-\mc{Q}_{(n+m)}\,  .
\ea\eeq
There are thus $4N$ conserved charges, $2N$ bosonic
and $2N$ fermionic. Only the $2N$ bosonic ones
are mutually commuting and independant.

Actually, it would seem at first sight that there are $(N+1)$
$\mc{I}_{(n)}$-type charges and thus a total of $2N+1$ mutually
commuting conserved charges. However,
it can be checked that say
$\mc{I}_{(N)}$ can be expressed in terms of the lower order
$\mc{I}_{(n)}$ as well as the
$\mc{H}_{(n)}$'s. Let us illustrate this in the simple context of
two particles ($N=2)$:
\begin{eqnarray}
\mc{H}_1&=&D_1+D_2\, ,\cr \mc{H}_2&=&D_1^2+D_2^2\, , \cr
\mc{I}_0&=&\theta_1\theta_1^\dagger+\theta_2\theta_2^\dagger\, , \cr
\mc{I}_1&=&\theta_1\theta_1^\dagger D_1+\theta_2\theta_2^\dagger D_2\,
.
\end{eqnarray}
The operator $\mc{I}_2$ depends of the preceding four conserved
quantities.  Indeed, it is easily
checked that
\begin{equation}
\mc{I}_2 = \mc{I}_1 \mc{H}_1 - \frac{1}{2} \mc{I}_0 \left( \mc{H}_1^2-
\mc{H}_2 \right) \, .
\end{equation}
It is clear that there can be no more than
$2N$ mutually commuting
conserved charges
since this is the maximal number of commuting operators in the free
case ($\beta=0$), in which case these quantities are
$\{p_i,
\theta_i\theta_i^\dagger\}$.

The srCMS model is thus seen to be integrable
in the usual sense.\footnote{The integrability can
also be established via the Lax formalism \cite{DLM1}. The form of
the Lax operator is actually the same as in the
non-supersymmetric case, except that every factor $X_{ij}$
(differentiated or not) is replaced by  $X_{ij}\kappa_{ij}$.}

\subsection{The superintegrability of the rsCMS model }

Consider the following $2N$ quantities:
\begin{eqnarray}
\mc{L}_n&=&\sum_{j}x_jD_j^{n+1}\, \Big|_{P^{S_N}}\, , \quad
\phantom{\theta_j\partial_{\theta_j}} n=-1,\ldots,N-2,\cr
    \mc{M}_n&=&\sum_{j}x_j\theta_j\partial_{\theta_j}D_j^{n+1}\,
\Big|_{P^{S_N}}\, ,\quad n=-1,\ldots,N-2 \, .
\end{eqnarray}
They satisfy the algebra
    \begin{equation}\label{vir}
    [\mc{L}_n,\mc{L}_m]=(n-m)\mc{L}_{n+m}\, ,\qquad
    [\mc{H}_n, \mc{L}_m]=n\mc{H}_{n+m} \, ,\qquad
    [\mc{H}_n,\mc{M}_m]=n\mc{I}_{n+m}\, .
\end{equation}
These relations are easily obtained using: \beq
[D_i^n,x_j]=\delta_{ij}\left(nD_i^{n-1}-\beta\sum_{k\neq
i}\frac{D_i^n-D_k^n}{D_i-D_k}K_{ij}\right)-\beta(1-\delta_{ij})\frac{D_i^n-D_j^n}{D_i-D_j}K_{ij}
\, , \eeq for $n=0,1,2\ldots$ and:\footnote{The expression of
$[x_iD_i^n,x_jD_j^m]$ contains many misprints in
\cite{Kuznetsov}.} \beqa
[\,x_iD_i^n,\,x_jD_j^m]&=&\delta_{ij}\left[(n-m)D_i^{n+m-1}+\beta
x_i\sum_{k\neq i}\frac{D_i^nD_k^m-D_i^m
D_k^n}{D_i-D_k}K_{ij}\right]\cr&&-\beta(1-\delta_{ij})\left[\frac{x_iD_i^{n+m}+x_jD_j^{n+m}}{D_i-D_j}K_{ij}
-(x_i+x_j)\frac{D_i^mD_j^n}{D_i-D_j}K_{ij}\right] \, , \eeqa for
$n,m=1,2,3\ldots$

We can construct from these,  $2N-2$ new and independent conserved
charges (i.e., commuting with  $\mathcal{H}_2$) :
\begin{eqnarray}
\mc{J}_n&=&\mc{H}_{n+1}\mc{L}_{-1}-\mc{L}_{n-1}\mc{H}_1\, ,\cr
\mc{K}_n&=&\mc{I}_{n+1}\mc{M}_{-1}-\mc{M}_{n-1}\mc{I}_1\, .
\end{eqnarray}
This a direct supersymmetric extension of the argument given in
\cite{Kuznetsov}\footnote{Furthermore, the algebra (\ref{vir})
obviously implies the
algebraic  linearization of the $4N$ equations of motion:
$$
\frac{d\mathcal{H}_n}{dt}=0
,\quad\frac{d\mathcal{I}_n}{dt}=0,\quad\frac{d\mathcal{L}_n}{dt}=-i\mathcal{H}_{n+2},\quad\frac{d\mathcal{M}_n}{dt}=-i\mathcal{I}_{n+2}\,
.$$ Such a linearization is also possible for the stCMS model.
It follows from  a direct generalization of the approach of
\cite{CFSa} (using the Lax formalism). On the other hand, note that
the superintegrability of the classical
rCMS model was proved in \cite{Wojciechowski:my}. }. The srCMS
model is thus not only {\it super} and {\it integrable}, but
also {\it superintegrable}.

\section{The stCMS model: construction of the eigenfunctions}

\subsection{The stCMS Hamiltonian}
  We now move to the second part of this work and discuss the
construction of the eigenfunctions of the sCMS models. If
the integrability structure of the (s)CMS models is most simply
analyzed in the rational case, the study of the
eigenfunctions is most naturally done in the trigonometric
case.\footnote{Of course there are no bound states in
the rational case. The presence of bound states  necessitates the
introduction of an harmonic confinement. But then the
eigenfunctions turn out (somewhat surprisingly) to be expressible in
terms of the eigenfunctions of the trigonometric case,
the Jack polynomials \cite{Sog}.}  The Hamiltonian of the stCMS model reads
\cite{SriramShastry:1993cz, DLM1}:
     \begin{equation}
\mc{H}_{(t)}=-\frac{1}{2}\sum_{i=1}^{N}
\partial_{x_i}^2+\left(\frac{\pi}{L}\right)^2\sum_{
i<j}\frac{\beta(\beta-\kappa_{ij} )}{\sin^2(\pi
x_{ij}/L)}-\left(\frac{\pi
\beta}{L}\right)^2\frac{N(N^2-1)}{6},
\end{equation}
where   $L$ is the circumference of the circle on which the particles
are confined.
Removing  the contribution of the
ground-state wave function,
\begin{equation}
\psi_0(x) =\Delta^\beta(x) \equiv\prod_{j<k}\sin^\beta\left(\frac{\pi
x_{jk}}{L}\right) \, ,
\end{equation}
the transformed Hamiltonian then
becomes
\begin{equation}
\bar{\mc{H}}\equiv
    \frac{1}{2} \left(\frac{L}{\pi}\right)^2
\Delta^{-\beta}\mc{H}_{(t)}\Delta^{\beta}
\, .
\end{equation}
Expressed in terms of the new bosonic variables
    $ z_j=e^{2\pi i x_j/L} $, it finally reads
\begin{equation} \label{shjack}
\bar{\mc{H}}= \sum_i (z_i \partial_i)^2+\beta \sum_{i<j}\frac{ z_i+z_j}
{z_{ij}}(z_i \partial_i-z_j\partial_j)-2\beta\sum_{i<j}\frac{z_i
z_j}{z_{ij}^2}(1-\kappa_{ij}) \, .
\end{equation}

\subsection{The $\bar{\mc{H}}$ eigenfunctions as symmetric superpolynomials}

The eigenfunctions of
the Hamiltonian (\ref{shjack})  are  superpolynomials, namely
polynomials of the bosonic variables $z_i$ and the fermionic ones $\theta_i$.
Looking for the proper generalization of the Jack polynomials, we
are interested in  eigenfunctions that are symmetric with
respect to the simultaneous interchange of both types of variables,
i.e., invariant under the action of $\mc{K}_{ij}$.

   Manifestly,
$\bar{\mc{H}}$ leaves invariant the space
of polynomials of a given degree in $z$ and a given degree in
$\theta$, being homogeneous in both sets of variables.
We consider eigenfunctions
of the  form:
\begin{equation}
\mc{A}^{(m)} (z,\theta;\beta)=\sum_{1 \leq i_1<i_2< \ldots <i_m \leq
N}\theta_{i_1}\cdots\theta_{i_m} A^{(i_1\ldots i_m)}(z;\beta)\, ,
\qquad m=0,1,2,3,\dots \, ,
\end{equation}
where $ A^{(i_1\ldots i_m)}$  is a homogeneous polynomial in  $ z$.
Due to the presence of  $ m $  fermionic variables in its  expansion,
    $ \mc{A}^{(m)} $  is said to belong to the  $ m$-fermion sector.
We stress that the simple dependence upon the
fermionic variables, which factorizes in  monomial prefactors, is a
consequence of their anticommuting nature.

We now clarify the symmetry properties, with respect to the $z$
variables, of the
eigenfunctions $\mc{A}^{(m)}$, assumed to be invariant under the action of the
   exchange operators
$\mc{K}_{ij}$.
Given that the $\theta$ products are antisymmetric, \emph{i.e.},
\beq
\kappa_{i_a i_b} \theta_{i_1}\cdots\theta_{i_m}=-
\theta_{i_1}\cdots\theta_{i_m} \quad\qquad\mbox{ if } \quad i_a,  i_b
\in
\{i_1,\ldots, i_m\} \, ,
\eeq
the superpolynomials $ A^{i_1\ldots i_m}$ must  be
partially antisymmetric to ensure the
complete symmetry of
$\mc{A}^{(m)}$.  More precisely, the functions
$ A^{i_1\ldots i_m}$ must satisfy the following
relations:\footnote{Note that the case
$m=1$, with $\mc{A}^{(1)} =\sum_{i}\theta_i A^{i}(z;1/\beta)$, is
special in that regard:
$K_{ij}A^{k}=A^{k}$ if and only if $ i,j \neq k$.}
\beq
\ba{l}
K_{ij}A^{i_1\ldots i_m}(z;\beta)=-
A^{i_1\ldots i_m}(z;\beta) \quad \forall \quad i\mbox{ and }j \in
\{i_1\ldots i_m\}\, ,\\
   K_{ij}A^{i_1\ldots i_m}(z;\beta)=\phantom{-}
A^{i_1\ldots i_m}(z;1/\beta) \quad \forall \quad i\mbox{ and }j
\not\in \{i_1\ldots i_m\}\, .
\ea
\eeq
We have thus established that any symmetric eigenfunction of the stCMS model
contains terms of {\it  mixed symmetry in} $z$:  each polynomial
$ A^{i_1\ldots i_m}$ is completely antisymmetric in the variables
$\{ z_{i_1},\ldots, z_{i_m}\}$, and totally symmetric in the
remaining variables
$z/\{z_{i_1},\ldots, z_{i_m}\}$ \cite{DLM1}.

To proceed further, we need to address the following points:  how to label
the eigenfunctions;  how to define a natural
basis for the space of superpolynomials;  how to define the
eigenfunctions via a triangular  expansion in that basis.
These points are considered successively in the following subsections.

\subsection{Superpartitions}

Symmetric
polynomials are indexed by
partitions.  In the same manner, symmetric
superpolynomials, i.e., polynomials in $P^{S_N}$, can be indexed by
{\it superpartitions}. A
superpartition in the $m$-fermion sector is a sequence of non-negative integers
that generates two partitions separated by a semicolon \cite{DLM1}: \beq
\Lambda=(\Lambda_1,\ldots,\Lambda_m;\Lambda_{m+1},\ldots,\Lambda_N)=
(\lambda^a ; \lambda^s), \eeq the first one being associated to an
antisymmetric function of the variables $\{ z_{i_1},\ldots, z_{i_m}\}$
\beq
\lambda^a=(\Lambda_1,\ldots,\Lambda_m), \qquad
\Lambda_i>\Lambda_{i+1} \qquad   i=1, \ldots m-1,\\
\eeq
and the second one, to a symmetric function of the variables $\{
z_{i_{m+1}},\ldots, z_{i_N}\}$
\beq
\lambda^s= (\Lambda_{m+1},\ldots,\Lambda_N), \qquad
\Lambda_i \ge \Lambda_{i+1} \qquad i=m+1, \ldots N.\\
\eeq
In the zero-fermion sector ($m=0$), the semicolon disappears and we recover the
partition $\lambda^s$.

   The weight (or degree) of a superpartition is
simply
the sum of its parts. For instance, the only possible
superpartitions of weight 3 in the 1-fermion sector are: \beq
(3;0), \quad (2;1),\quad (1;2), \quad (1;1,1), \quad (0;3), \quad
(0;2,1), \quad (0;1,1,1),
\eeq while in the 2-fermion sector, they are:
\beq (3,0;0) \quad (2,1;0)\quad (2,0;1)\quad (1,0;2)\quad (1,0;1,1). \eeq


\subsection{The monomial symmetric superpolynomials  basis }

Given our goal of constructing the superextension of the Jack
polynomials, which themselves decompose triangularly in the
symmetric monomial basis, the superextension of the latter will
provide our natural expansion basis.
The monomial symmetric superpolynomials (supermonomials for short)
are defined as \cite{DLM1}:
\begin{equation}
m_{\Lambda}(z,\theta)=\sm{\Lambda_1,\ldots,
\Lambda_m;\Lambda_{m+1},\ldots,\Lambda_{N}}(z,\theta)={\sum_{\sigma\in
S_{N}}}' \theta^{\sigma(1, \ldots, m)}z^{\sigma(\Lambda)}, \end{equation}
where the prime
indicates that the  summation is
restricted to distinct terms, and where
\begin{equation}
z^{\sigma(\Lambda)}=z_1^{\Lambda_{\sigma(1)}} \cdots z_m^{\Lambda_{\sigma(m)}}
z_{m+1}^{\Lambda_{\sigma(m+1)}} \cdots
z_{N}^{\Lambda_{\sigma(N)}}  \quad {\rm{and}} \quad
\theta^{\sigma(1, \ldots, m)} = \theta_{\sigma(1)} \cdots
\theta_{\sigma(m)} \, .
\end{equation}
Clearly, in the zero-fermion sector, a supermonomial reduces to an ordinary
symmetric monomial.\footnote{There are other simple bases for the
space of symmetric superpolynomials. For instance,
one can combine the $2N$ algebraically independent symmetric power
sums,
$$p_n=\sum_iz_i^n=m_{(;n)},\qquad q_{n-1}=\sum _i \theta_i z_i^{n-1}=m_{(n;0)} ,\qquad n=1,\ldots,N, $$ to form a
new basis: $$p_\Lambda=q_{\lambda^a}
p_{\lambda^s}=q_{\Lambda_1}\cdots
q_{\Lambda_m}p_{\Lambda_{m+1}}\cdots p_{\Lambda_N}\, .$$ One can
also generate the  whole space with the elementary symmetric
superfunctions, $$e_n=\sum_{1\leq i_1<\ldots<i_n\leq
N}z_{i_1}\cdots z_{i_n}=m_{(;1^n)},\qquad f_{n-1}=\sum_{1\leq
j\neq i_1,\ldots,i_{n-1}\leq N\atop 1\leq i_1<\ldots<i_{n-1}\leq
N}\theta_j z_{i_1}\cdots z_{i_{n-1}}=m_{(0;1^{n-1})},$$ or with
the complete symmetric superfunctions,
$$h_n=\sum_{1\leq i_1\leq\ldots\leq i_n\leq
N}z_{i_1}\cdots z_{i_n}=\sum_{\Lambda\atop
|\Lambda|=n,\overline{\underline{\Lambda}}=0}m_\Lambda,\qquad
j_{n-1}=\sum_{\Lambda \atop
|\Lambda|=n-1,\overline{\underline{\Lambda}}=1}(\Lambda_1+1)\,m_\Lambda,$$
where $n=1,\cdots,N$ while $|\Lambda|=\sum_i\Lambda_i$ and
$\underline{\overline{\Lambda}}=m$ stand for the degrees in $z$
and $\theta$ respectively.} Here is an example with $N=4$ that
neatly illustrates the mixed symmetry of each component:
\begin{eqnarray}
m_{(1,0;1,1)}&=& \theta_1 \theta_2 (z_1-z_2)(z_3 z_4)+\theta_1
\theta_3 (z_1-z_3)(z_2 z_4)
\nonumber\\& +&\theta_1 \theta_4 (z_1-z_4)(z_2 z_3)+\theta_2 \theta_3
(z_2-z_3)(z_1 z_4)  \nonumber\\ & +&\theta_2 \theta_4
(z_2-z_4)(z_1 z_3)+\theta_3 \theta_4 (z_3-z_4)(z_1 z_2)
\end{eqnarray}
Two other examples (with $N=3$) will clarify the different role of a
zero entry in each  sector:
\begin{eqnarray}
   m_{(3;0)}&= &  \theta_1 z_1^3+ \theta_2 z_2^3 + \theta_3 z_3^3  \nonumber\\
   m_{(0;3)}&= &\theta_1 (z_2^3+z_3^3)+ \theta_2 (z_1^3+z_3^3)+
\theta_3 (z_1^3+z_2^3) \, .
\end{eqnarray}

\subsection{Jack superpolynomials: a first definition}\label{sjackmon}

We now define our first candidates for the role of
Jack  superpolynomials (in the $m$-fermion
sector) as the
unique eigenfunctions of the supersymmetric Hamiltonian
   $\bar{\mc{H}}$,
\begin{equation}
\bar{\mc{H}}\, \mc{J}_\Lambda (z,\theta;\beta)  =
\varepsilon_\Lambda\mc{J}_{\Lambda}(z,\theta;\beta)\, ,
\end{equation}
   that can be decomposed triangularly in terms of monomial
superfunctions:
\begin{equation}
\mc{J}_\Lambda (z,\theta;\beta)=m_{\Lambda} (z,\theta)+\sum_{\Omega;\, \Omega
<  \Lambda}c_{\Lambda,\Omega}(\beta) m_{\Omega}(z,\theta)\, .
\end{equation}
Clearly, for this definition to be complete, we need to specify the
ordering ($<$)  underlying the triangular decomposition. The
simplest and most natural choice at this point is to formulate the
ordering in terms of the partitions $\Lambda^*$
and
$\Omega^*$ associated respectively to the  superpartitions $\Lambda$
and $\Omega$,
   by rearranging their parts in decreasing order.
(For instance, the
rearrangement
of $\Lambda=(7,4,3,1;8,6,5,3,3,1)$ is the partition $\Lambda^*
=(8,7,6,5,4,3,3,3,1,1)$.)
We thus say that $\Omega<\Lambda$ if
$\Omega^*<\Lambda^*$ with respect to the dominance ordering on partitions:
\begin{equation}
        \Lambda^*\geq \Omega^*\mtext{iff}
       \Lambda_1^*+\Lambda_2^*+\dots
+
\Lambda_i^* \geq \Omega_1^* + \Omega_2^* +\dots +\Omega_i^* \, ,
\quad \forall i \, ,
\end{equation}
with $\Lambda_i^*= (\Lambda^*)_i$.
This turns out to be sufficient to calculate explicitly the Jack
superpolynomials  \cite{DLM1}. Here is an example of such a polynomial:
\begin{equation}
\mc{J}_{(2;2)}=
\sm{2;2}+\frac{2\beta}{1+\beta}\sm{2;1^2}+\frac{\beta}{1+\beta}\sm{1;2,1}+\frac{6\beta^2}{(1+\beta)(1+2\beta)}\sm{1;1^3}
\, .
\end{equation}

Although complete, the above characterization is not quite precise.
Indeed, not all supermonomials $m_\Omega$
labeled by a superpartition $\Omega$ dominated by $\Lambda$ in the
above way do  appear in the decomposition. For
instance, the terms $m_{(0;2,1,1)}$ and $ m_{(0;1^4)}$, although
allowed by the ordering condition, are not present in the
expression of
$\mc{J}_{(2;2)}$. This poses the following natural question: can we
characterize {\it precisely} the terms that appear in
the supermonomial triangular decomposition of
$\mc{J}_\Lambda$, that is, can we pinpoint the precise ordering at work?
A second (and more ambitious) related question is the
following: can we write down an explicit formula for the coefficients
$c_{\Lambda,\Omega}$?

The answer to both of these questions is {\it yes}. The clue to obtain the
answer  lies in the following
observation: the action of
$\bar{\mc{H}}$ on the supermonomial basis is triangular and this
triangularity determines precisely the ordering entering in
the definition of the eigenfunctions. Moreover, the action of
$\bar{\mc{H}}$ can be computed exactly, essentially because
it can be reduced to a two-body computation. This turns out to
provide all the data required for evaluating the coefficients
$c_{\Lambda,\Omega}$. These conclusions are completely general and
apply to any CMS model, supersymmetric or not.  In the
next subsections, we indicate the key steps in reaching these
conclusions, keeping the presentation
rather general.  Technical details can be found in \cite{DLM2}.

\subsection{Generalities: the triangular action of the Hamiltonian,
the induced  ordering and a determinantal
formula for the eigenfunctions}

Consider a generic Hamiltonian $H$ and a generic basis $m_a$.
(Typically, the $m_a$ will be  monomial functions but another
basis could be used.) Suppose that, in a given subspace\footnote{In
the context of the stCMS, this subspace could be fixed by
the set of superpartitions
with given degree $n$ and given fermion number $m$. However, this
could also be a subspace pertaining to a a
non-supersymmetric problem, hence characterized by some conditions on
ordinary partitions.} spanned by three basis elements
$m_a,\, m_b ,\,m_c$ (for some labels
$a,b,c)$, the action of
$H$ takes the following form:
\begin{eqnarray}\label{hsurm}
Hm_{a}&= & \epsilon_a m_{a}+v_{ab}m_{b}+ v_{ac}m_{c}\nonumber \\
Hm_{b}&= & \epsilon_b m_{b}+v_{bc}m_{c} \nonumber\\
Hm_{c}&= & \epsilon_c m_{c} \, ,
\end{eqnarray}
where the eigenvalues are all distinct.
The action of $H$ is obviously triangular. It is
also clear that the triangularity entails an ordering:
$a>b>c.$ The key point is that this ordering is precisely the one
governing the triangular expansion of the eigenfunctions of
$H$,
denoted $J_a$, in the $m_a$ basis.  This follows from the following
result: if $H$ acts on the $m_a$'s as in (\ref{hsurm}),
its eigenfunction reads \cite{LLM}:
\begin{equation}\label{jadet}
J_a\propto\left| \begin{array}{ccc}
m_c&m_b&m_a\cr
\epsilon_c-\epsilon_a & v_{bc}&v_{ac}\cr
0&\epsilon_b-\epsilon_a & v_{ab}\end{array}\right| \, .
\end{equation}
Let us verify that $(H-\epsilon_a)J_a$ does indeed vanish. Since the
entries of the determinant are numbers except for the
first row, $H$ acts nontrivially only on this row:
\begin{equation}
(H-\epsilon_a)J_a\propto
\left| \begin{array}{ccc}
(\epsilon_c-\epsilon_a)m_c&(\epsilon_b-\epsilon_a)m_b+v_{bc}m_{c}&(\epsilon_a-\epsilon_a)m_a+v_{ab}m_{b}+

v_{ac}m_{c}\cr
\epsilon_c-\epsilon_a & v_{bc}&v_{ac}\cr
0&\epsilon_b-\epsilon_a & v_{ab}\end{array}\right| \, .
\end{equation}
The coefficient of $m_a$ is identically zero. The coefficient of
$m_b$ is most simply obtained by setting $m_c=0$. This leads
to
\begin{equation}
(H-\epsilon_a)J_a\big|_{m_c=0}\propto
\left| \begin{array}{ccc}
0&(\epsilon_b-\epsilon_a)m_b&v_{ab}m_{b}\cr
\epsilon_c-\epsilon_a & v_{bc}&v_{ac}\cr
0&\epsilon_b-\epsilon_a & v_{ab}\end{array}\right| \, .
\end{equation}
The first row being proportional to the third one,  the determinant
is  zero. Similarly, by
setting $m_b=0$, we see that the first two rows become proportional,
which again enforces the vanishing of the determinant.
Since the expansion coefficients of $(H-\epsilon_a)J_a$ in the
monomial basis are all zero, we can conclude that
$(H-\epsilon_a)J_a=0$.

The eigenfunction $J_a$ is thus given by the determinant
(\ref{jadet}) up to a multiplicative constant.  Note that it is not equal
to zero
  because the eigenvalues being all distinct, the coefficient
of $m_a$ cannot vanish.   But this shows readily
(i.e., by expanding the determinant) that it can be written under the
form $J_a\propto(m_a+\cdots)$ where the dots refer to
lower order terms in  {\it the ordering induced by the action of $H$
in the $m_a$ basis.} This is precisely the point we
wanted to emphasize.

Note that the eigenfunction can be determined uniquely by simply
enforcing its leading expansion coefficient to be one (monic
condition).

This simple example shows neatly that the ordering governing the
decomposition of $J_a$ in the $m_a$ basis is encoded in
the triangularity of $H$ on $\{m_a\}$. However, it is somewhat
misleading in its simplicity. It suggests that all terms
occurring in the decomposition of $J_a$ can be compared with each
others and that a single `chain of ordering' (which
refers to the case where all $m_i$ with $i$ comparable to $a$ do
appear in the action of $H\,m_a$) is always involved.
We will thus consider a second slightly more complicated case that
captures the generic features.

Consider a five-dimensional subspace spanned by
$m_a,\, m_b ,\,m_c,\,m_d,\,m_e$ and suppose the following  action of
Hamiltonian:
\begin{eqnarray}
Hm_{a}&= & \epsilon_a m_{a}+v_{ab}m_{b}+ v_{ac}m_{c}+ v_{ad}m_{d} \nonumber\\
Hm_{b}&= & \epsilon_b m_{b}+v_{bc}m_{c} \nonumber\\
Hm_{c}&= & \epsilon_c m_{c} +v_{ce}m_{e}\nonumber\\
Hm_{d}&= & \epsilon_d m_{d} +v_{de}m_{e}\nonumber\\
Hm_{e}&= & \epsilon_e m_{e} \, .
\end{eqnarray}
This action implies the following chains of ordering: $a>b>c$ as well as
$c>e$ and $d>e$. Hence, we see that the ordering within
the subspace is only {\it partial} because for instance $c$ and $d$
cannot be compared (this is also the case for $b$ and
$d$). The corresponding eigenfunction is\footnote{The validity of
this result requires that eigenvalues, corresponding to labels
that can be compared, be distinct.}
\begin{equation}
J_a\propto\left| \begin{array}{ccccc}
m_e&m_d&m_c&m_b&m_a\cr
\epsilon_e-\epsilon_a & v_{de}&v_{ce}&0&0\cr
0&\epsilon_d-\epsilon_a & 0&0&v_{ad}\cr
0&0&\epsilon_c-\epsilon_a & v_{bc}&v_{ac}\cr
0&0&0&\epsilon_c-\epsilon_a & v_{ab}
\end{array}\right| \, .
\end{equation}
It thus follows that both $m_c$ and $m_d$ appear in the expression of
$J_a$, even if $c$ and $d$ cannot be compared
with each other. The main point is that they are both comparable to $a$.
This example also illustrates our second point:
$e$ is comparable to $a$ but $m_e$ does not appear in the
expression of $H\,m_a$. In other words, $e$ is compared to $a$ by
a sequence of two chains of orderings:
$a>b>c$ and
$c>e$. We will later relate the number of chains of ordering and the
number of applications of a ladder operator on
labels.

The construction of determinantal expressions for the eigenfunctions
provides a
strong motivation for obtaining explicitly the coefficients
$v_{ab}$ appearing in the action of $H$ on the monomial basis: they
are the building blocks of the expansion
coefficients
$c_{ab}$ of $J_a$ in that basis.  The knowledge of  these
coefficients leads thus to {\it closed-form expressions for  the
eigenfunctions!}  Quite remarkably, these coefficients are indeed
computable, as we will now show.

\subsection{Generalities: the explicit action of $H$ by the universal
dressing  of a model-dependent $N=2$
computation}

As we just noticed, the specification of the ordering requires the
determination of the coefficients $v_{ab}$ that do not vanish in the
   action of $H$ on $m_{a}$.  Actually, we will see that it is not much
more complicated to compute precisely all these
coefficients
$v_{ab}$. And, as pointed out, this yields directly the eigenfunctions.

The strategy is the
following: we first compute the action of $H$ in the two-particle
sector. Such computations are always very easy. For
a CMS-type model, whose Hamiltonian is a sum of two-body
interactions, this computation turns out to encode the core value of
the coefficient
$v_{ab}$. Indeed, to go from
$N=2$ to a general
$N$ simply amounts to dressing the result by a symmetry factor
\cite{Sogo, LLM, DLM2}. And another remarkable fact is that,
although the
$N=2$ computation is model
dependent, {\it the symmetry dressing appears to be universal}, that
is, the symmetry factor boils down to a symmetry of the
labels (partitions or superpartitions) of the eigenfunctions, hence
independent of the root structure, or the rational or
trigonometric version of the (s)CMS under consideration.  This is
certainly so for all the cases we have considered so far
(including the rational case with confinement \cite{DLM4}).

An immediate objection could be formulated with regards to this
program: although it is clear that the action of $H$ on $m_a$
is well-defined in the $N=2$ sector, we know that for a general
$N$-body problem, the action of $H$ on a particular element
$m_a$, that itself does not vanish upon reduction to $N=2$, may contain
terms that disappear upon reduction. Are these terms
properly taken into account?

To make the above considerations more concrete and precise,  we
will now return to the stCMS model. This will also allow us to
address the potential objections in a definite context.

\subsection{The transposition of the action of $\bar{\mc{H}}$ on
superpartitions: ladder operators}

   When the action of the Hamiltonian is a sum of two-body terms $\sum
\bar{\mc{H}}_{ij}$ (up to a derivative part that acts
diagonally), as is the case for the (s)tCMS model,   the various
terms appearing in the decomposition of $\bar{\mc{H}}
m_\La$ can be characterized by the action of a ladder operator
$R_{ij}^{(\ell)}$ acting on the superpartitions $\La$.\footnote{If in
the Hamiltonian, there are in addition some terms
that act non-diagonally (as in the (s)rCMS
model \cite{DLM4}), their action is taken into account by another
  ladder operator  acting on a single entry of a
superpartition.} More precisely, $R_{ij}^{(\ell)}$ acts on two parts
$\La_i$ and
$\La_j$ of a superpartition as \cite{DLM2}:
\begin{equation} \label{rij}
R_{ij}^{(\ell)}(\Lambda_1,\dots,\Lambda_i,\dots,\Lambda_j,\dots,\Lambda_{N})
=\left\{
\begin{array}{ll}
(\Lambda_1,\dots,\Lambda_i-\ell,\dots,\Lambda_j+\ell,\dots,\Lambda_{N}) &
{\rm{if }} \, \, \Lambda_i > \Lambda_j\,, \cr
(\Lambda_1,\dots,\Lambda_i+\ell,\dots,\Lambda_j-\ell,\dots,\Lambda_{N}) &
{\rm{if }} \, \, \Lambda_j > \Lambda_i\,.
\end{array}\right.
\end{equation}
for  $ i<j $  and  $ \ell \geq 0 $.
This action of  $ R_{ij}^{(\ell)} $  is non-zero only in the
following cases:
\begin{equation} \label{rijl}
\begin{array}{rl}
{\rm{I}}: & \quad i,j \in \{1,\dots,m \}\,  \qquad {\rm{and}} \qquad
\lfloor \frac{\Lambda_i-\Lambda_j-1}{2}\rfloor \geq \ell \,,\cr
{\rm{II}}: & \quad i \in \{1,\dots,m\} \, , j \in \{m+1,\dots,N \}\,
\qquad {\rm{and}} \qquad
|\Lambda_i-\Lambda_j|-1 \geq   \ell\,,\cr
{\rm{III}}: & \quad i,j \in \{m+1,\dots,N \}\,  \qquad {\rm{and}} \qquad
\lfloor \frac{\Lambda_i-\Lambda_j}{2}\rfloor \geq \ell \,.\cr
\end{array}
\label{type}
\end{equation}
Here $\lfloor x \rfloor$ stands for
the largest integer smaller or equal to
$x$. This characterization is obtained by a consideration of the action of
$\bar{\mc{H}}$ in the $N=2$ sector. Three cases must
then be distinguished, according to their fermion number: $m=2,1,0$,
corresponding respectively to pairs $(i,j)$ of type
I, II, III. The results pertaining to ordinary Jack polynomials
correspond to case III.

   Let us make the induced ordering (denoted $\geq^s$) explicit.
Given a sequence
\begin{equation}
\gamma=(\gamma_1,\dots,\gamma_m;
\gamma_{m+1},\dots,\gamma_N)\,,
\end{equation} we will denote by $\overline{\gamma}$
the superpartition whose antisymmetric part is the
rearrangement of $(\gamma_1,\dots,\gamma_m)$ and whose symmetric part is
the rearrangement of $(\gamma_{m+1},\dots,\gamma_N)$.
For example, we have
\begin{equation}
\overline{(1,3,2;2,3,1,2)}
=(3,2,1;3,2,2,1).
\end{equation}
We say that
\begin{equation} \Lambda \geq^{s} \Omega\qquad {\rm  iff}\quad
\Omega =
\overline{R_{i_k,j_k}^{(\ell_k)} \dots \overline{R_{i_1,j_1}^{(\ell_1)}
\Lambda}}
\end{equation}
for a given sequence of operators
    $ R_{i_1,j_1}^{(\ell_1)},\dots,R_{i_k,j_k}^{(\ell_k)} $.
This is a refinement of the ordering introduced previously at the
level of the corresponding partitions $\La^*$ and a
generalization of the dominance ordering, which is recovered in the
zero-fermion sector.

All the supermonomials $m_\Omega$ appearing in the expansion of
$\bar{\mc{H}}\, m_\La$ are precisely those whose labeling
superpartition is obtained from $\La$ by {\it one application} of the
ladder operator $R$, that is, all $\Omega$ that can be
written as
$\Omega =
\overline{R_{i,j}^{(\ell)}
\Lambda}$ for some $i<j$, and $\ell>0$. This implies that the
number of non-zero parts of $\Omega$ exceeds that of $\La$ by
at most one. On the other hand, the supermonomials
$m_\Gamma$ appearing  in the expansion of $\mc{J}_\La$ (which is now
understood to represent a
determinant expansion) are
precisely those whose label
$\Gamma$ is obtained from $\La$ by {\it any number of applications}
of the ladder operator $R$. We can now make the connection
with the loose terminology of the previous sections: one application
of $R$ corresponds to a single chain of ordering while
multiple chains of ordering  are described by the action of a sequence
of ladder operators.

The expressions for the coefficients $v_{\La,\Omega}$, as well as a
detailed description of the symmetry factors, can be
found in \cite{DLM2} and will not be repeated here. Let us instead
address the potential problem pointed out previously:
the action of $\bar{\mc{H}}$ on a two-part
superpartition may contain more than two parts, hence lie beyond
the $N=2$ sector, which was claimed to contain all the relevant
information. Consider for instance
\begin{equation}
\bar {\cal{H}}\, m_{(3;1)}= (10+4 \, \beta \,  N -6\, \beta) \, m_{(3;1)} +
    2 \, \beta\, m_{(2;2)}
+ 8 \, \beta \, m_{(2;1,1)}+2 \, \beta \, m_{(1;2,1)}
\end{equation}
which holds for arbitrary values of $N>2$.
However, if we specialize to $N=2$, the last two terms disappear. How
could these be taken into account by a $N=2$
computation? The point is that the different terms $m_\Omega$ in
$\bar {\cal{H}}\, m_\La$ are not necessarily linked to the {\it same}
two-body calculation. While the contribution of the first non-diagonal term
$ m_{(2;2)}$ can be described by the two-body interaction
$\bar {\cal{H}}_{12}$, that is,
$(2;2)=R_{12}^{(1)} (3;1)$, this is not the proper two-body
interaction for the description of the other two terms. Indeed,
we see that $ (2;1,1)= R_{13}^{(1)} (3;1,0)$. This shows that the
relevant $N=2$ computation is rather $\bar
{\cal{H}}_{13}m_{(3;*,0)}$. Similarly, because $(1;1,2)= R_{13}^{(2)}
(3;1,0)$ and $\overline{(1;1,2)}= (1;2,1)$, the
corresponding two-body problem is again $\bar {\cal{H}}_{13}m_{(3;*,0)}$.

It should be heavily stressed that the remarkable fact that the
result for general $N$ can be deduced out of the $N=2$ one is
true for the action of
$\bar {\cal{H}}$ in the supermonomial basis; this would not be
possible directly for the eigenfunctions $\mc{J}_\La$.

\subsection{$\mc{J}_\La$ as eigenfunctions of the $\bar{\cal{H}}_n$ charges}

The method sketched in the previous subsections indicate that we can
construct closed-form expressions for the
eigenfunctions  $\mc{J}_\La$ of the stCMS Hamiltonian $\bar
{\cal{H}}$, our first candidates for the role of
Jack superpolynomials. Actually, we can
prove that the $\mc{J}_\La$'s diagonalize the whole tower of
conserved charges $\bar {\cal{H}}_n$ \cite{DLM3} (recall that the
bar indicates that the contribution of the groud state wave function
has been taken away):
\begin{equation}
\bar {\cal{H}}_n \mc{J}_\La= \epsilon_\La^{(n)}\mc{J}_\La \, .
\end{equation}
Unfortunately these superpolynomials are {\it not orthogonal}. In
other words, the action of the $\bar {\cal{H}}_n$'s
leaves a residual degeneracy. Indeed, we can check that
$\epsilon_\La^{(n)}=\epsilon_\Omega^{(n)} $ for all $n$  if
$\La^*=\Omega^*$ even when
$\La\not=\Omega
$ \cite{DLM3}.  However, in retrospect, this is not too surprising:
we have not constructed the common eigenfunctions
of {\it all} the commuting conserved charges. We still need to diagonalize the
$\bar{\cal{I}}_n$ charges.

\subsection{Constructing orthogonal eigenfunctions: diagonalization of the
$\bar{\cal{I}}_n$ charges}

As just indicated, the  eigenfunctions of the Hamiltonian
$\bar{\cal{H}}= \bar{\cal{H}}_2$ happen to be eigenfunctions of
all the
$\bar{\cal{H}}_n$ operators. Similarly, to study of the  eigenfunctions
of the complete set   $\{\bar{\cal{I}}_n\}$, it
suffices to consider only the first non-trivial charge $\bar{\cal{I}}_1$
\cite{DLM3}.

The remaining problem is thus to construct linear combinations of the
$\mc{J}_\La$ that are eigenfunctions of
$\bar{\cal{I}}_1$. The strategy is by now clear: we first compute the
action of $\bar{\cal{I}}_1$ on the $\mc{J}_\La$
basis, determine the new ordering induced by the underlying
triangularity, compute the expansion coefficients exactly, and
then write down the eigenfunctions, now denoted $J_\La$, in
determinantal form. The self-adjointness property of the charges
$\{\bar{\cal{H}}_n\}$ and $\{\bar{\cal{I}}_n\}$ ensures the
orthogonality of the $J_\La$. These are thus the {\it genuine
Jack superpolynomials}.  The details  of this construction can be
found in {\cite{DLM3}.

Let us mention that a more direct way to build the orthogonal
Jack superpolynomials
is also presented in {\cite{DLM3}.   It amounts to
simply symmetrizing the product of a non-symmetric
Jack polynomial with a monomial in the fermionic variables.
Even though this approach is more direct, we prefer not to include it
since it is not as much in the
spirit of this
presentation.

\subsection{A remark on the underlying Lie superalgebraic structure}

Let us conclude this section on the stCMS model by addressing the
following question:  is
there a Lie superalgebra structure underlying this problem?   Given that the
tCMS Hamiltonian is closely linked to
the
$su(N)$ root structure, the natural  guess is that the stCMS Hamiltonian
would be related to the superalgebra $su(m,N-m)$. The root
structure of $su(m,N-m)$ is (see for instance \cite{Dico}):
\begin{equation}
\delta_i-\delta_j \, ,\quad \epsilon_k-\epsilon_\ell\, ,\quad
\delta_i-\epsilon_k \, ,\quad \quad 1\leq i<j\leq
m\quad{\rm and} \quad  m+1\leq k<\ell\leq N-m
\end{equation}
with
\begin{equation}
\delta_i\cdot\delta_j = -\delta_{ij}\, ,\qquad
\epsilon_k\cdot\epsilon_\ell= \delta_{k\ell}\, ,\qquad
\delta_i\cdot\epsilon_k=0
\end{equation}
When $m=0$, we recover the $su(N)$ roots. The roots
$\delta_i-\delta_j$ and $\epsilon_k-\epsilon_\ell$ are said to be
bosonic while the remaining roots $\delta_i-\epsilon_k$ are called
fermionic. However, it should be clear that there is no
genuine notion of statistics at the level of the roots themselves
(although the generators do have a
definite statistics) since for instance the difference between two
fermionic roots could be bosonic
(e.g.,
$\delta_1-\epsilon_1-(\delta_2-\epsilon_1)=\delta_1-\delta_2$). This,
of course, is not a deep observation. However, it
readily implies that the Hamiltonian constructed from the root system
of a Lie superalgebra, $su(m,N-m)$ for instance, does
not contain anticommuting variables (see \cite{Serg} for example).
Hence,  the stCMS  Hamiltonian does not appear to have an
immediate interpretation in terms of the $su(m,N-m)$ algebra.

However, there does exist a link with the $su(m,N-m)$ root structure,
but at the level of the action of the ladder operator
$R_{ij}^{(\ell)}$. Referring to equations (\ref{rij}) and
(\ref{rijl}), we see that the action of $R_{ij}^{(\ell)}$ in case I
and III corresponds to the subtraction of positive bosonic roots,
while in case II, it amounts to subtract a positive or a
negative fermionic root.  That both positive and negative fermionic
roots are involved is not suprising given the relativity of
the positivity requirement for these roots. At the level of
superpartitions, the sign of the fermionic roots should not play
any role because it is simply related to our
choice of relative ordering  for the two partitions $\lambda^a$
and $\lambda^s$ composing the superpartition.

\section{Conclusion}

The integrability and superintegrability of the sCMS models rely
on a fundamental observation:  the supersymmetric
extension is fully captured by a fermionic exchange operator. This
allowed us to infer the (super)integrability by means of a
projection argument. Many generalizations of the srCMS model could
be shown, in a similar way, to be superintegrable: the srCMS
models formulated for general root systems, the extension of the
rCMS model with many supersymmetries, the srCMS model with spin
degrees of freedom, etc.

On the other hand, we have stressed that the method outlined in section 3 for
constructing eigenfunctions is quite general. It has already been
extended to the srCMS case with confinement, that is, to the
construction of orthogonal generalized  Hermite (super)polynomials
\cite{DLM4}. Generalized Jacobi and Laguerre polynomials in
superspace could also be obtained along this line from the
sr/tCMS model with $B$-type roots. The method  can be easily
extended to models with more supersymmetries or including spin
degrees of freedom. The astonishing power of the method gives us
hope that this line of attack could provide a breakthrough in the
study the eigenfunctions of
elliptic models.

    \vskip0.3cm
\noindent {\bf ACKNOWLEDGMENTS}

We thank F. Lesage for a useful remark. P.D. would like to thank
the Fondation J.A Vincent for a student fellowship and the work of
L.L. and P.M. is supported by NSERC.

\end{document}